\documentclass[aps,pra,reprint]{revtex4-2}

\usepackage[mathlines]{lineno}
\usepackage{blindtext}
\usepackage[export]{adjustbox}
\usepackage{graphicx,epsfig,epstopdf}
\usepackage{amsmath,amssymb}
\usepackage{esvect}
\usepackage{mathtools}
\usepackage{color}
\usepackage{subfigure}
\usepackage{array}
\usepackage{tabularx}
\usepackage{multirow}
\usepackage{booktabs}
\usepackage{mathtools}
\usepackage{float}
\usepackage{comment}
\newcolumntype{L}[1]{>{\raggedright\arraybackslash}p{#1}}
\newcolumntype{C}[1]{>{\centering\arraybackslash}p{#1}}
\newcolumntype{M}[1]{>{\centering\arraybackslash}m{#1}}
\allowdisplaybreaks

\def\_#1{{\bf #1}}

\def\.{\cdot}

\def\H0{{H_0}}
\def\E0{\eta_0 {H_0}}

\def\=#1{\overline{\overline #1}}

\begin{document}

\title{Time-modulated circuits and metasurfaces for emulating arbitrary transfer functions}
\author{G.~A.~Ptitcyn$^{1,2}$}
\author{M.~S.~Mirmoosa$^{1}$}
\author{S.~Hrabar$^{3}$}
\author{S.~A.~Tretyakov$^1$}

\affiliation{
$^1$Department of Electronics and Nanoengineering, Aalto University, P.O.~Box 15500, 
FI-00076 Aalto, Finland\\
$^2$Department of Electrical and Systems Engineering, University of Pennsylvania,
Philadelphia, PA 19104, U.S.A.\\
$^3$Faculty of Electrical Engineering and Computing, University of Zagreb, Unska 3, 10000 Zagreb, Croatia}

\begin{abstract}

Temporal modulation unlocks  possibilities to dynamically control and modify the response of electromagnetic systems. Employing explicit dependencies of circuit or surface parameters on time enables the engineering of systems with conventionally unachievable functionalities. Here, we propose a novel approach that enables the emulation of electromagnetic systems that can have arbitrary frequency dispersion and nonlinear properties, including the non-Foster response. In particular, we show that a proper modulation of a time-varying capacitor allows one to mimic a static inductance, capacitance, or resistance having arbitrary values, both positive and negative. We discuss necessary modifications of determined ideal modulation functions that ensure the stability of the system. To demonstrate the applicability of the proposed method, we introduce and simulate an invisible sensor, i.e., a device that does not produce any scattering and is capable of sensing. Three different geometries are proposed and validated using full-wave simulations. In addition to that, we discuss the stability of the systems that are modulated externally. We believe that this study introduces a new paradigm of using time modulations to engineer system responses that can be applied not only to  electromagnetic systems (in electronics, microwaves, and optics) but also to other branches of physics. 
\end{abstract}
\maketitle  


\section{Introduction} 
Unlike static electromagnetic systems, parameters of time-varying systems depend explicitly on time.
This dependence brings to another level the number of possibilities to control electromagnetic fields, e.g.~\cite{Engheta_4D}. For instance, one can design efficient systems that realize unconventional functionalities~\cite{Our2,Halevi,pacheco2020temporal,pacheco2021temporal,lustig2018topological}. In addition to that, many conventional limitations imposed on static systems can be overcome using time modulations, for instance, enhancing radiation from electrically small antennas beyond the Chu limit~\cite{li2019beyond,Mostafa_dipole,mostafa2022antenna}, breaking time-reversal symmetry and achieving magnetless nonreciprocity  limits~\cite{hadad2016breaking,yu2009complete,sounas2014angular,shi2017optical,dinc2017synchronized,fleury2018non,taravati2017nonreciprocal,Our2}, accumulation of energy with reactive elements~\cite{Our1}, versatile   absorption control~\cite{mostafa2022coherently}, and more. 

Introducing time modulation one can often get an original solution even for such basic and well understood problem as impedance matching.
The notion of impedance matching is used for waves of all nature, including acoustical and mechanical. Impedance matching in electromagnetics is a common problem at all frequency ranges, although sometimes it is called differently. Nevertheless, the conventional ways to tackle this problem, such as resonant matching, are well known, but usually they provide an ideal solution at a single frequency only. In the vicinity of this frequency, impedance mismatch is negligible, which makes it possible to efficiently use the system within this frequency range. Typically, associate relative bandwidth is few percent, outside of which impedance mismatch becomes huge and the device cannot be used. Alternatively to the resonant impedance matching, one could implement so-called ``non-Foster" impedance matching~\cite{sussman2009non}. Non-Foster elements do not obey Foster's theorem that states that the driving-point  impedance  of a lossless one-port network must be a positive-real and odd function of the Laplace-transformed frequency variable $s$~\cite{foster1924reactance}. Consequently, the corresponding reactance must monotonically increase with the frequency~\cite{sussman2009non}. Typical non-Foster reactive elements are negative capacitors and negative inductors. These devices are of active nature and their dispersion is an inverse of the dispersion of ordinary reactive elements with positive parameters. Therefore, non-Foster elements can compensate for the dispersion of ordinary elements, thus providing broadband operation of various metamaterials, metasurfaces, and antenna structures~\cite{hrabar2010towards,hrabar2011negative,saadat2012composite,chen2013broadening,white2012non}. 
 
Non-Foster elements are primarily realized using negative impedance converters (NICs) that were introduced by Latour~\cite{latour_1928}, Dolmage~\cite{DOLMAGE1932} and~Mathes~\cite{Mathes1930} and developed by Linvill later in 1953~\cite{linvill1953transistor}. Since then this concept was adopted for realization of many applications: impedance matching devices~\cite{sussman2009non,white2012non}, epsilon-near-zero metamaterials~\cite{hrabar2011negative,hrabar2013ultra}, broadband cloaks~\cite{hrabar2010towards,chen2013broadening}, broadband and efficient amplifiers~\cite{akwuruoha201764,akwuruoha201855}, broadband phase shifters~\cite{lee20156,al2018wide}. Essentially, non-Foster elements are active electronic circuits that mimic the required load impedances using a positive feedback loop. Unfortunately, the use of positive feedback very often leads  to inherent instability.
On the other hand, it has been shown recently that time modulation provides an alternative route for the realization of non-Foster circuit elements~\cite{hrabar2020time}.

In this paper, we show how one can mimic not only non-Foster elements but any static, stationary (non-dispersive), time-varying, and even nonstationary (dispersive) systems using a single time-varying circuit element. We show several practical examples that demonstrate the flexibility of the proposed method, for instance, we show how one can mimic capacitance, inductance, and resistance of any value (including negative) using only a time-varying capacitor. In addition to that we study the proposed time-varying circuit elements in application to sensors. Modulation of a capacitive layer enables to hide of the whole device, producing no scattering. Three different designs of the structure are proposed. Finally, we analyze the stability of systems with the proposed time-varying circuit elements and find that they are stable under the assumption of external modulation. 





\section{Emulation of arbitrary dispersive networks} 

Let us consider a dispersive one-port network that is causal, linear, and time-invariant (LTI). This network is described by a transfer function which is called impedance or admittance depending on the excitation source. The relation between the electric current $i(t)$ and the voltage $v(t)$ corresponding to the LTI network is given by (e.g., Refs.~\cite{Landau_Lifshitz,ptitcyntutorial})
\begin{equation}
i(t)=\int\limits_0^{\infty}\tilde{Y}(\gamma)v(t-\gamma)\mathrm{d}\gamma,
\label{eq:LTINY}
\end{equation}
where $\tilde{Y}(\gamma)$ is the admittance kernel, and $\gamma$ and $t$ represent the retardation and observation times, respectively. Note that in the above definition we assume that the network is excited by a voltage source. 

Our goal is to find a single time-varying reactive component that would provide the same output $i(t)$ when it is excited by the same voltage source $v(t)$. The reactive component can be a capacitor or an inductor. In this work, we choose a capacitor with the capacitance $C(t)$. Under the assumption of the instantaneous response of the materials from which the capacitor is made, the electric current flowing through a time-varying  capacitor is expressed as
\begin{equation}
i(t)=\frac{\mathrm{d}}{\mathrm{d}t}\bigg[C(t)v(t)\bigg].
\label{eq_time_varying_capac_IR}
\end{equation}
It is worth mentioning that one may modify this equation for the electric current taking into account arbitrary dispersion of time-varying capacitors~\cite{ptitcyntutorial}. However, here, for simplicity let us continue with the instantaneous response assumption. 

We require the electric currents defined by Eqs.~\eqref{eq:LTINY} and \eqref{eq_time_varying_capac_IR} to be equal at all times and find the corresponding temporal function for the capacitance. After a simple algebraic manipulation, we deduce that time-modulation of capacitance according to 
\begin{equation}
C(t)=\frac{1}{v(t)}\bigg(\beta+\int\int_0^{\infty}\tilde{Y}(\gamma)v(t-\gamma)\mathrm{d}\gamma\mathrm{d}t\bigg)
\label{eq:FLTI}
\end{equation}
ensures that the current through this capacitor is the same as given by \eqref{eq:LTINY}. 
Here, we have introduced a constant $\beta$ which can be chosen arbitrarily. However, there are practical limitations. First, as the above equation shows, to ensure that the function $C(t)$ has no singularities, the voltage $v(t)$ should never become zero. This means that $v(t)$ has to be DC-biased if it is varying in time as a time-harmonic signal. Another practical concern is imposed by the fact that $C(t)$ needs to be practically realizable, and, hence, it cannot be negative. Thus, it is required that $C(t)>0$ for all $t$. The importance of the choice of constant $\beta$ emerges at this point. In fact, this requirement is ensured by properly adjusting $\beta$. 

Before moving further, we stress that the derivation of Eq.~\eqref{eq:FLTI} was done for mimicking an LTI network. However, the network can be nonlinear. The key point is that the nonlocality relation between the output and the input related to the network should be definitely changed. Indeed, following the fundamentals of nonlinear optics~\cite{Landau_Lifshitz}, if the nonlinearity is weak, we can rewrite the current-voltage relation as
\begin{equation}
\begin{split}
i(t)&=\int\limits_{0}^{\infty}\tilde{Y}_1(\gamma_1)v(t-\gamma_1)\mathrm{d}\gamma_1\cr
&+\iint\limits_{0}^{\infty}\tilde{Y}_2(\gamma_1,\gamma_2)v(t-\gamma_1)v(t-\gamma_2)\mathrm{d}\gamma_1\mathrm{d}\gamma_2,
\end{split}
\end{equation}
in which $\tilde{Y}_1$ and $\tilde{Y}_2$ characterize the network. By equating the above equation with Eq.~\eqref{eq_time_varying_capac_IR}, we find the corresponding temporal function of the capacitance. Intriguingly, we mimic the characteristics of a nonlinear network by using a nondispersive time-varying capacitor which is a linear component. This paves the road to realizing interesting functionalities usually provided due to nonlinearity by employing temporal modulations of linear systems.

\section{Emulation of arbitrary circuit elements} 

In the following, we make our theoretical investigation narrower and suppose that the LTI network is composed of individual bulk circuit elements. First, we show that time-varying capacitances can mimic a static capacitance $C_{\rm eq}$ with  arbitrary values including negative ones. For that, we specialize the general approach that is outlined above. In order to find the appropriate modulation function for the variable capacitance, we need to equate the expressions for electric currents flowing in the two circuits: With the desired equivalent capacitance $C_{\rm eq}$ and with the time-varying capacitance $C(t)$. Regarding the equivalent capacitance, the electric current is given by
\begin{equation}
i(t)=C_{\rm eq}\frac{\mathrm{d}v(t)}{\mathrm{d}t},
\label{eq_negative_C_charac}
\end{equation}
where we assume that $C_{\rm eq}$ can have an arbitrary value. On the other hand, for the case of a time-varying capacitance, we employ Eq.~\eqref{eq_time_varying_capac_IR} and \eqref{eq_negative_C_charac}. The result reads
\begin{equation}
C(t)=\frac{c_1}{v(t)}+C_{\rm{eq}},
\label{eq_mod_function_for_C}
\end{equation} 
in which $c_1$ is a constant. However, the above equation can be also directly derived based on the general expression given by Eq.~\eqref{eq:FLTI}. Indeed, in Eq.~\eqref{eq:FLTI}, the admittance kernel is the inverse Fourier transform of $Y(\omega)=j\omega C_{\rm{eq}}$, in which ``$j$" is the imaginary unit, and $\omega$ denotes the angular frequency. Using the Fourier transform properties, we see that the admittance kernel is the time derivative of the Dirac delta function multiplied by the equivalent capacitance, i.e., $\tilde{Y}(\gamma)=C_{\rm{eq}}\delta'(\gamma)$. By substituting this admittance kernel into Eq.~\eqref{eq:FLTI}, and by remembering the following feature of the Dirac delta function: 
\begin{equation*}
\int f(x)\delta^{n}(x)\mathrm{d}x=-\int\frac{\mathrm{d}f(x)}{\mathrm{d}x}\delta^{n-1}(x)\mathrm{d}x,
\end{equation*}
and, also, recalling that $\mathrm{d}f(y-x)/\mathrm{d}x=-\mathrm{d}f(y-x)/\mathrm{d}y$, we successfully achieve the same result as expressed by Eq.~\eqref{eq_mod_function_for_C}.

As mentioned before, there are practical limitations in using Eq.~\eqref{eq_mod_function_for_C}. Suppose that the voltage is a time-harmonic signal  $v(t)=V_{\rm{DC}}+v_{\rm{ac}}\cos(\omega_0t+\phi)$. In order to have a nonzero denominator at all moments of time, we see that the DC bias voltage $V_{\rm{DC}}$ needs to be larger than $v_{\rm{ac}}$. In addition, if $c_1$ is a positive number, we readily show that $c_1>-(V_{\rm{DC}}+v_{\rm{ac}})C_{\rm{eq}}$. Using this condition, we properly choose the value of $c_1$ so that the time-varying capacitance never becomes negative.  

Next, we extend the analysis and show how time-varying capacitances can be also employed to enticingly emulate static resistances and inductances. In the case of linear resistance, the voltage-current relation simply reads $i(t)=v(t)/R_{\rm eq}$, where $R_{\rm eq}$ is the equivalent resistance. By comparing it with the current flowing through a time-varying capacitance, one can derive the corresponding modulation function for $C(t)$ as
\begin{equation}
C(t)=\frac{c_2}{v(t)}+\frac{1}{R_{\rm eq}v(t)}\int v(t)\mathrm{d}t.
\label{eq_equivalentR}
\end{equation} 
This expression confirms that it is possible to provide the functionality of virtual absorption in electromagnetic systems by using time-varying capacitors~\cite{Our1}. Probably, the only challenge is that for time-harmonic sources, the corresponding voltage should be DC-biased to avoid singularities. Because there is  voltage integration in Eq.~\eqref{eq_equivalentR}, such DC component of voltage gives rise to a function that increases linearly in time.

To realize inductive response, we need to be cautious. Capacitors and inductors respond quite differently to a DC source. While the capacitance is open-circuited resulting in  zero electric current, the inductance is oppositely short-circuited corresponding to  zero voltage. To overcome this issue, we consider a lossy inductance, which is represented by the resistance $R_{L}$ connected in series with a lossless inductance $L_{\rm eq}$. In this case, the relation between the voltage and current is expressed as
\begin{equation}
v(t)=L_{\rm eq}\frac{\mathrm{d}i(t)}{\mathrm{d}t}+R_Li(t).
\label{eq_voltage_RLeq}
\end{equation}
One can mimic such a circuit with a lossy time-varying capacitance, where loss is characterized by the resistance $R_C$ connected in parallel to $C(t)$. The electric current flowing in this circuit is written as 
\begin{equation}
i(t)=\frac{\mathrm{d}}{\mathrm{d}t}\bigg[C(t)v(t)\bigg]+\frac{v(t)}{R_C}.
\label{eq_current_CtR}
\end{equation} 
Again, we require equivalence of the above two equations, Eqs.~\eqref{eq_voltage_RLeq} and \eqref{eq_current_CtR}, to find the corresponding function for time modulation of the capacitance. However, in this case the solution is  not so straightforward. First, we integrate both parts of Eq.~\eqref{eq_voltage_RLeq} with respect to time and express the integral of current as
\begin{equation}
\int i(t)\mathrm{d}t=-\frac{L_{\rm eq}}{R_L}i(t)+\frac{c_1}{R_L}+\frac{\int v(t)\mathrm{d}t}{R_L},
\label{eq_RLeq_intcur}
\end{equation}
where $c_1$ is the integration constant. As the next step, we repeat the same integration regarding Eq.~\eqref{eq_current_CtR}. Thus, we have
\begin{equation}
\int i(t)\mathrm{d}t=C(t)v(t)-c_2+\frac{1}{R_C}\int v(t)\mathrm{d}t
\label{eq_current_CtR1},
\end{equation}
in which $c_2$ is another integration constant. Equations \eqref{eq_RLeq_intcur} and \eqref{eq_current_CtR1} must be identical. Consequently, we obtain an expression for $C(t)$:
\begin{equation}
C(t)=\frac{1}{v(t)}\bigg[\frac{c_1}{R_L}+c_2\bigg]-\frac{L_{\rm eq}}{R_L}\frac{i(t)}{v(t)}+\bigg[\frac{1}{R_L}-\frac{1}{R_C}\bigg]\frac{\int v(t)\mathrm{d}t}{v(t)}.
\label{eq_C_mimicking_L}
\end{equation}
For emulating static capacitances and resistances, the information about voltage over the element was enough for determining the temporal modulation function for  capacitance (see Eqs.~\eqref{eq_mod_function_for_C} and \eqref{eq_equivalentR}). In comparison, here, we clearly observe that in addition to voltage, information about the electric current is also necessary. Notice that in actual implementations resistances $R_L$ and $R_C$ cannot be zero, as shown by the equation, however, they can be both negative or have the opposite signs. Interestingly, in the scenario when $R_L=R_C$, the last term of Eq.~\eqref{eq_C_mimicking_L} vanishes, and the modulation function becomes an oscillating function (since there is no integration of the voltage or current) which makes it convenient for practical realizations.

We emphasize that the interested reader can follow the logic presented in this section and derive a dual set of equations for a time-varying inductance, although from a practical point of view, it may be less favorable. 

 \begin{figure*}
\centering
\includegraphics[width=0.9\textwidth]{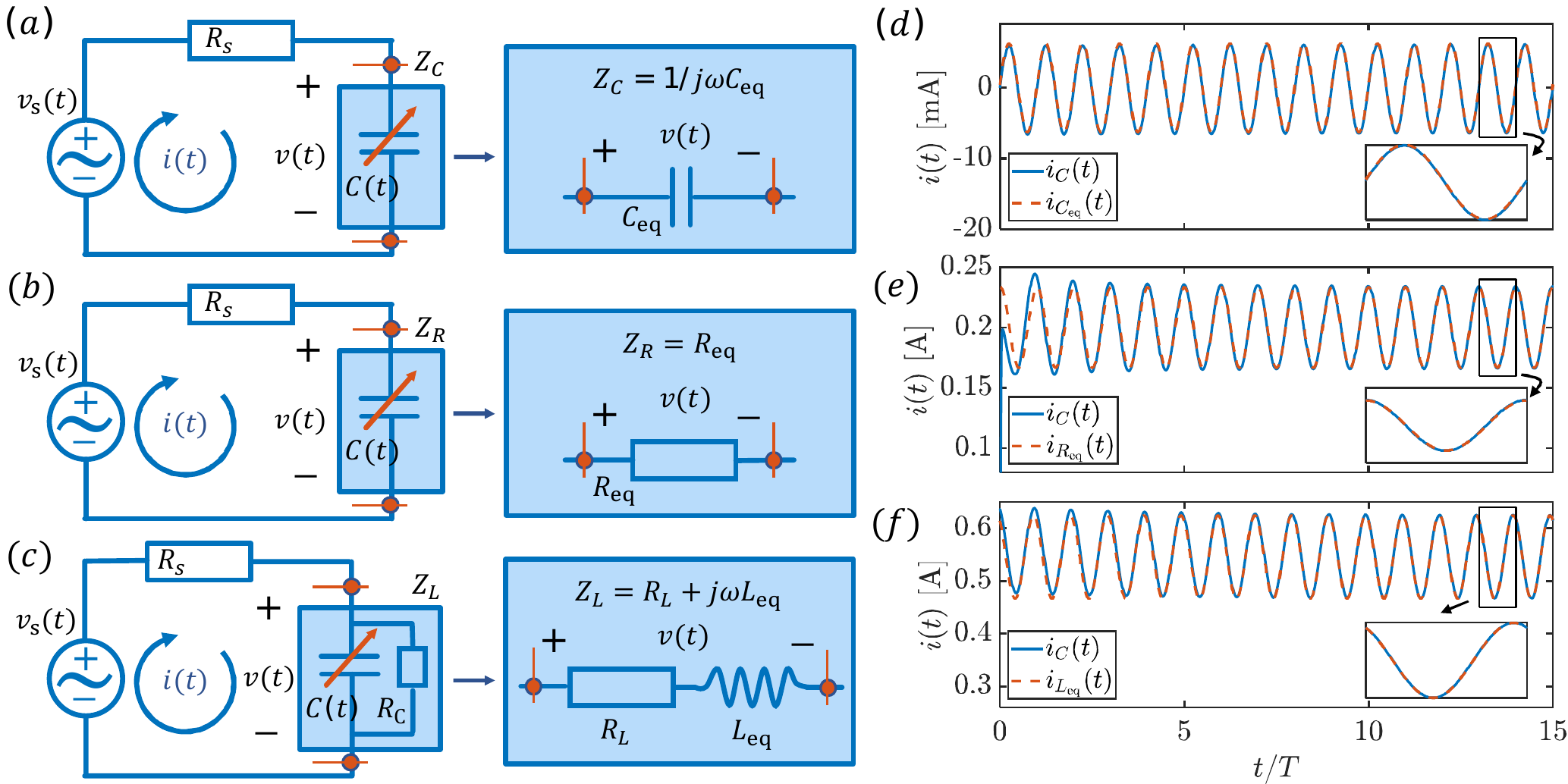}
\caption{(a)--Electric circuit with a time-varying capacitance that can mimic a capacitor with an arbitrary $C_{\rm eq}$. (b)--Electric circuit with a lossy time-varying capacitance that mimics a lossy inductor  with an arbitrary value of $L_{\rm eq}$. (c)--Electric circuit with a time-varying capacitance that mimics $R_{\rm eq}$ with an arbitrary value. (d), (e), and~(f)--Currents flowing through circuits in (a), (b),~and~(c). Currents flowing through the equivalent elements $C_{\rm eq}$, $R_{\rm eq}$, and $L_{\rm eq}$ are calculated in the steady-state regime.}
\label{fig1_mimicking_arbitraryLandC}
\end{figure*}


\section{Mimicking non-Foster elements and arbitrary resistances}

Equations~\eqref{eq_mod_function_for_C}, \eqref{eq_equivalentR},  and \eqref{eq_C_mimicking_L} present expressions that allow mimicking static circuit elements by using a time-varying capacitor $C(t)$. Controlling free parameters of these equations, one always obtains a realizable modulation profile for $C(t)$, which makes this approach quite appealing for practical implementations. Interestingly, one can even mimic 
non-Foster static reactive elements (i.e., $C_{\rm eq}$ and $L_{\rm eq}$ whose values are smaller than zero) and negative resistance as well ($R_{\rm eq}<0$). 

However, realizing static non-Foster elements may result in instabilities. 
In fact, linear time-invariant systems with ideal non-Foster components connected to ordinary positive-parameter elements are always unstable. Let us briefly demonstrate this feature by giving an example of a circuit formed by an excitation source and a positive resistance connected in series with an ideal capacitance $C_0$, whose value is frequency-independent and negative. This simple electric circuit is governed  by a 
first-order linear differential equation. The complementary solution of this equation which gives the transient response is easily calculated as 
$v_{\rm{h}}(t)=a_1\exp(-{t}/{RC_0})$, where $a_1$ is a constant defined by the initial conditions. According to this relation, it is evident that the voltage across an ideal non-Foster capacitor ($C_0<0$) grows exponentially in time. Now, let us recall that the time modulation of $C(t)$ accordingly to \eqref{eq_mod_function_for_C}  makes the current flowing in the 
time-varying capacitor identical to the current through the equivalent element. Therefore, the two circuits including the time-varying capacitor and the equivalent element should share the same properties. Consequently, since the static circuit including the non-Foster element is unstable, the equivalent circuit with the time-varying capacitor is also unstable. 

In known realizations of non-Foster elements (based on the use of active circuits with appropriate feedback circuits), the issue of potential instability is handled by making the active device dispersive, so that it provides gain only in a certain limited frequency range. In this case, the equivalent negative capacitor exhibits negative capacitances only in a limited range of frequencies. In the proposed time-modulation approach, similarly, it is necessary to make the emulation of a negative capacitor not perfect, avoiding transition-time instabilities.   One efficient method is to completely damp the transient response (corresponding to the complementary solution) by replacing the instantaneous voltage $v(t)$ in the definition of the modulation function by the required steady-state voltage. Practically, this means that instead of a feedback modulation device that modulates the capacitor according to the actual voltage across is, we modulate externally by a voltage source giving the expected steady-stage voltage over a negative-capacitance circuit. Alternatively, it is possible to add a low-pass filter to the instantaneous values of voltage $v(t)$ in the device that modulates $C(t)$.  This way, effectively, we limit the operating bandwidth of the device, removing the inherent instability of ideal non-Foster elements. 

In the following, the ideal modulation functions in Eqs.~\eqref{eq_mod_function_for_C}, \eqref{eq_equivalentR}, and \eqref{eq_C_mimicking_L} are modified to avoid instabilities. The ideal voltage $v(t)$ and current $i(t)$ are replaced with $v_{\rm cap}(t)$ and $i_{\rm cap}(t)$, respectively. The new voltage and current represent the ideal ones with a truncated spectrum.  In order to prove the feasibility of the introduced approach, we consider three different circuits shown in 
Fig.~\ref{fig1_mimicking_arbitraryLandC} (see (a)--(c)). In this figure, panels (d)--(f) show the simulation results performed for these corresponding circuits in Simulink. For all three circuits, the source voltage is $v_{\rm s}(t)=6+\cos{\omega t}$~V, where the radial frequency is $\omega=2\pi\times 1$~MHz (therefore, the period is $T=1\,\mu$sec). The source resistance is chosen to be $R_{\rm s}=10~\Omega$. The equivalent capacitance that is synthesized by $C(t)$ in the example of Fig.~\ref{fig1_mimicking_arbitraryLandC}(a) is $C_{\rm eq}=-1$~nF. The equivalent resistance synthesized in the example of Fig.~\ref{fig1_mimicking_arbitraryLandC}(b) is $R_{\rm eq}=10~\Omega$. Finally, for the case of arbitrary inductance in Fig.~\ref{fig1_mimicking_arbitraryLandC}(c), its equivalent value is chosen to be $L_{\rm eq}=-1~\mu$H with the loss resistance $R_L=1~\Omega$ connected in series. The electric currents flowing through ideal equivalent elements are calculated in the steady-state regime assuming impedances $Z_C=1/j\omega C_{\rm eq}$, $Z_{R}=R_{\rm eq}$, and $Z_L=R_L+j\omega L_{\rm eq}$. In Figs.~\ref{fig1_mimicking_arbitraryLandC}(d--f), we explicitly observe that for all cases both obtained electric currents, the one corresponding to the time-varying capacitor and the one corresponding to the equivalent element, are in good agreement after a transition period.

Therefore, the conceptual solutions given by  Eqs.~\eqref{eq_mod_function_for_C}, \eqref{eq_equivalentR}, and \eqref{eq_C_mimicking_L} define stable circuits with appropriately modified modified modulation functions. This approach can be used, for example, for broadband impedance matching of small antennas and in principle can be applied in more sophisticated structures such as distributed  transmission-line circuits and metasurfaces.

\section{Invisible time-modulated metasurface as a sensor}
\begin{figure*}
\centering
\includegraphics[width=0.9\textwidth]{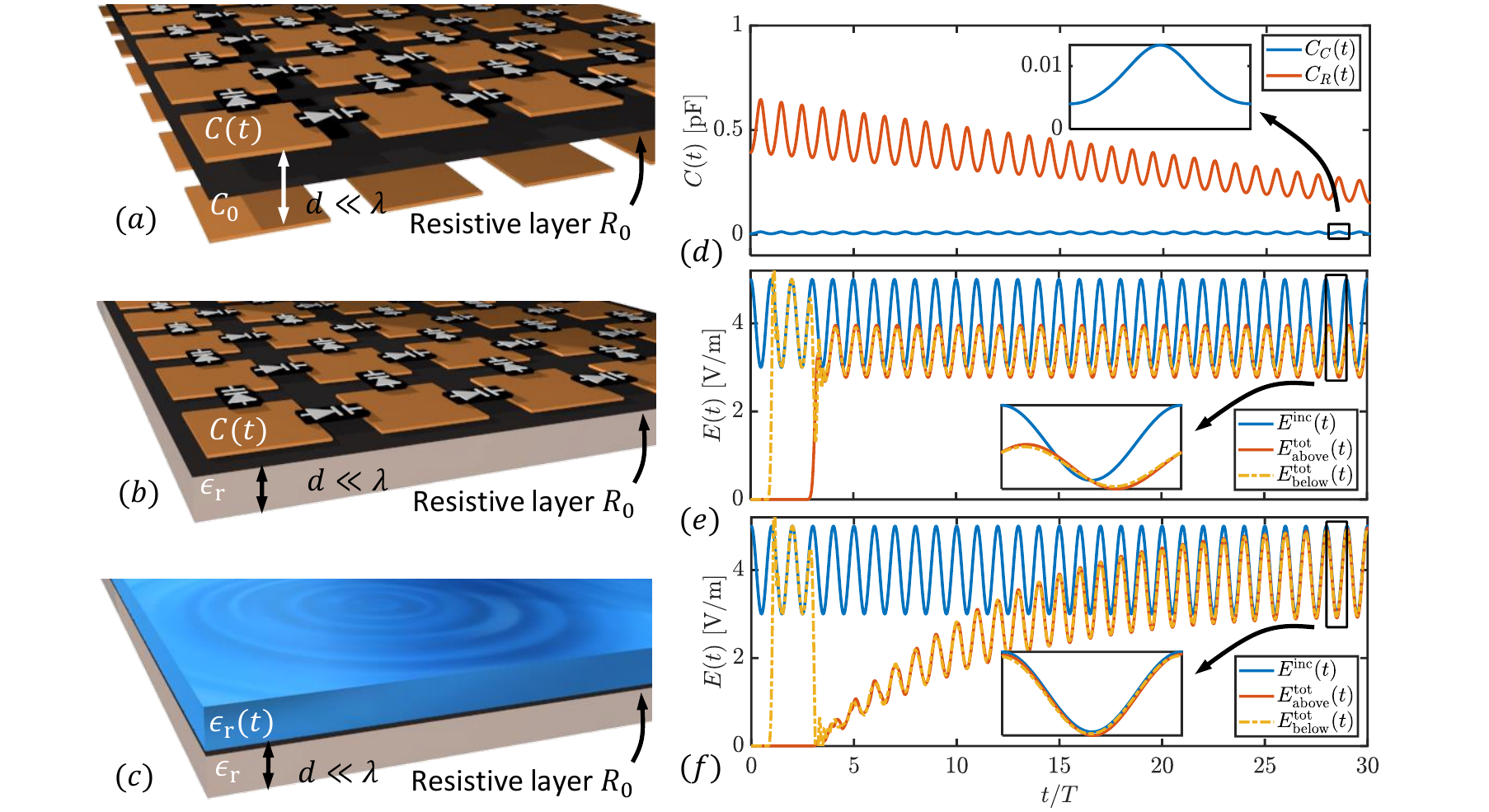}
\caption{(a)--Concept of an invisible sensor made of two arrays of patches, separated by a small distance $d$, and a resistive layer $R_0$ placed in the middle. The capacitance of one of the arrays is modulated in time using varactors. (b)--Conceptual realization of an invisible sensor using an array of patches placed on top of a thin dielectric substrate. Varactors in the gaps between the patches realize variation of the layer properties in time. A thin resistive sheet $R_0$ is placed in between the capacitive layer and the substrate. (c)--Concept of an invisible sensor made of two thin dielectric layers and a resistive sheet $R_0$ in between. The permittivity of one of the layers is modulated in time. (d)~and~(e)--Electric field at distance $\lambda$ above and below the metasurface compared with the incident field in the absence of modulation~(d) and in the case when modulation is turned on~(e).}
\label{fig_invisible_sensor}
\end{figure*}

A very peculiar response appears when a non-Foster capacitive sheet is placed close to a usual one. The non-Foster counterpart of the usual capacitive sheet creates an electric current density of equal strength that flows in the opposite direction making the total current density zero. Practically, it means that the structure is fully transparent for external excitation at a given frequency although the current densities in each of the sheets do not vanish. This property can be exploited for sensing applications. However, for this purpose, absorption is essential. Hence, we locate a resistive sheet between the two reactive ones. In the general case, this device would inevitably cast a shadow, due to the presence of loss. To address this issue, one can adjust the modulation function that would account for this loss of energy, thus realizing an invisible sensor. This can be done simply by combining two modulation functions: One for emulating negative surface capacitance and another one for emulating negative surface resistance. Accordingly, the virtual negative surface resistance results in generating energy that is exactly equal to the dissipated energy (i.e., the energy extracted from the external field). Let us demonstrate how such device can be realized.

Figure~\ref{fig_invisible_sensor}(a--c) shows various possible realizations of a device in form of thin electric-current sheets. All these designs are essentially equivalent to a parallel connection of a static capacitance $C_0=10$~fF, resistance $R_0=1000~\Omega$, and time-varying capacitance $C(t)$. The structure is excited by a plane wave $E_{\rm in}(t)=E_{\rm DC}+E_0\sin(\omega t)$, where $E_{\rm DC}=4$~V/m, $E_0=1$~V/m, and the radial frequency is $\omega=2\pi\times100$~GHz. In order to cancel the effect of the static elements $C_0$ and $R_0$, the variable capacitance $C(t)$ needs to include two functions $C_C(t)$ and $C_R(t)$ that emulate the static elements with the opposite sign, i.e., $-C_0$ and $-R_0$. For the function  $C_C(t)$, we can assume external modulation and replace the ideal voltage $v(t)$ in Eq.~\eqref{eq_mod_function_for_C} with the modified voltage elucidated in the previous section. Therefore, the modulation function is expressed as 
\begin{equation}
C_C(t)=\frac{c_1}{E_{\rm in}(t)}-C_0,
\label{eq_invisible_modulation}
\end{equation}
where $c_1=C_0\times7$~V/m, which in principle can be an arbitrary number that ensures $C_C(t)>0$ at all $t$. It is worth noting that Eq.~\eqref{eq_invisible_modulation} contains exactly $E_{\rm in}(t)$ in the denominator since it represents the total field on the device in the steady-state regime. In the case of invisibility, from the equivalent circuit model point of view, the total resistance and the total capacitance of the device vanish. As a result, there is no reflection at all, and this essentially means that the total field is equal to the incident field. Modulation function $C_R(t)$ is calculated by applying Eq.~\eqref{eq_equivalentR}, where the voltage $v(t)$ is substituted by $E_{\rm in}(t)$, and the equivalent resistance is obviously chosen as $-R_0$. Note that $C_R(t)$ as the modulation function for realizing negative resistance eventually becomes smaller than zero, which makes the system unstable. However, by adjusting the free parameter $c_2$, one can tune the moment when $C_R(t)$ crosses zero. Here, we chose $c_2=14 c_1$. 

Concerning the first two designs, the same time-varying capacitive layer $C(t)=C_C(t)+C_R(t)$ is used. The difference between the two designs is that the second one exploits equivalence of a thin dielectric substrate and a capacitive sheet with $C_{\rm eff}=(\epsilon_{\rm{r}}-1)d/(\eta_0c_0)$ ($\epsilon_{\rm{r}}$ and $d$ are the relative permittivity and the substrate thickness, respectively, $\eta_0$ represents the free-space intrinsic impedance, and $c_0$ denotes the speed of light). By choosing a dielectric substrate with thickness $d=\lambda/400$, where $\lambda$ is the wavelength of the incident wave, and the relative permittivity $\epsilon_{\rm r}=151.7$, one obtains $C_{\rm eff}=C_0$. For such a small thickness and large permittivity, the difference between a dielectric substrate and a capacitive sheet becomes negligible. The resistive layer $R_0$ is placed in between the substrate and the time-varying layer (see Fig.~\ref{fig_invisible_sensor}(b)). 

The third realization uses the same static dielectric substrate as in the second design, however, the time-varying capacitive sheet is replaced with a thin dielectric slab, whose relative permittivity is changing in time as 
\begin{equation}
\epsilon^{\rm m}_{\rm r}(t)=\bigg[C_C(t)+C_R(t)\bigg]\frac{\eta_0c_0}{d}+1,
\end{equation}
where $C_C(t)$ and $C_R(t)$ is the same as in the previous designs. The thickness of the time-varying layer is also equal to $d=\lambda/400$. Since all the designs are conceptually the same, they produce almost identical scattering (see Fig.~\ref{fig_suppl_Sens_iden_scat} in Appendix). Therefore, here, simulation results are shown for only the last design because it is the most challenging from a computational point of view. Figure~\ref{fig_invisible_sensor}(d) shows the modulation functions $C_C(t)$ and $C_R(t)$. Since the envelope of $C_R(t)$ linearly decays, we stop simulations before the modulation function becomes negative. Figure~\ref{fig_invisible_sensor}(e) indicates the incident and total electric fields above and below the suggested device at distance $\lambda$, for the case when modulation is turned off. In fact, these fields represent a reference. It is clear that static capacitive and resistive layers bring strong scattering, meaning that the amplitude and phase of the total fields change substantially. Figure~\ref{fig_invisible_sensor}(f) shows the case when modulation is turned on. It clearly demonstrates how time modulation makes the device invisible, with the total fields above and below the metasurface being identical to the incident field.

Notice that the device proposed in Fig.~\ref{fig_invisible_sensor}(a--c) can operate in a wide range of frequencies. The only restriction is the thickness of the device (it needs to remain much smaller than the operating wavelength $d\ll\lambda$). This peculiarity makes the proposed device broadband. Interestingly, for small absorption, which does not need to be hidden, one could use only $C_C(t)$ which is a harmonic function, and, thus, the implementation becomes much easier in practice. Accordingly, similar realization scenarios illustrated in this section can be introduced (see Fig.~\ref{fig_supl_not_a_Sensor} in Appendix with the analysis of these structures).


\section{On stability of the proposed time-modulated devices}

Here, we argue that temporal modulation of reactive elements for realizing non-Foster components result  in stable systems providing that the spectrum of the modulation voltage $v(t)$ is finite and function $v(t) $ has no singularities. Let us consider the case of emulating negative capacitance with a time-varying one, i.e., Eq.~\eqref{eq_mod_function_for_C}. The spectrum of voltage should contain only the frequency components within the desired operational frequency range. In particular, we consider a single-frequency operation and modulate the capacitor by the signal $v_{\rm cap}(t)$ on the capacitance, which is the fixed time-harmonic function at the frequency of operation, with a DC bias. In this case, the modulation is external, and the negative capacitance is properly realized only at the desired operational frequency. In practical realizations, this approach can be extended to variable inputs by realizing modulation as a feedback loop with a frequency filter in the device that modulates $C(t)$, leading to replacing of voltage $v(t)$ in Eq.~\eqref{eq_mod_function_for_C} by its frequency-filtered version $v_{\rm cap}(t)$. Although the modulation is not fixed anymore, we expect that transient instabilities are eliminated also in this case due to the limited bandwidth of negative capacitance.

\begin{figure}
\centering
\includegraphics[width=0.45\textwidth]{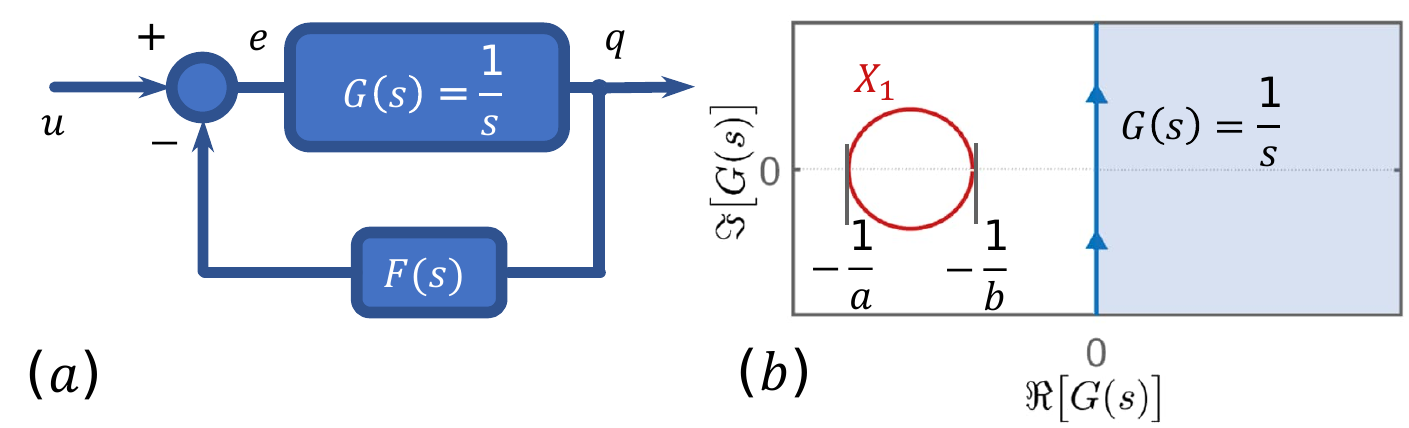}
\caption{Preliminary figure. (a)--Control systems equivalent to the studied system. (b)-- The Nyquist diagram of the system.}
\label{fig4_Nyquist_stability}
\end{figure}

To analyze the stability of the circuit shown in Fig.~\ref{fig1_mimicking_arbitraryLandC}(a), where $C(t)=c_1/v_{\rm cap}(t)+C_{\rm eq}$, we can use the Nyquist criterion for linear time-varying systems~[Sec.~8.5 in Ref.~\onlinecite{d1970linear}]. For the case of the considered simple circuit, the differential equation for charge reads:
\begin{equation}
\frac{\mathrm{d}q(t)}{\mathrm{d}t}+\frac{1}{RC(t)}q(t)=\frac{1}{R}v_{\rm s}(t),
\label{eq_RCt_charge}
\end{equation}
where $q(t)$ is the charge on the capacitance. The equivalent feedback system for the circuit under study is presented in Fig.~\ref{fig4_Nyquist_stability}(a). First, we introduce the transfer function $G(s)$ that has a similar physical meaning as in the ``ordinary" Nyquist approach: It corresponds to the differential equation \eqref{eq_RCt_charge} without the time-modulated term. In the Laplace domain, function $F(s)$ characterizes external time-varying feedback signal, i.e., the modulation function. A particular system is stable if the locus of the function $G(s)$ does not intersect with the critical circle $X_1$ defined by the bounds of the modulation function. 

The transfer function in the Laplace domain $G(s)$, in this case, is simply $1/s$, whose locus occupies the positive (right half) of the complex plane (see Fig.~\ref{fig4_Nyquist_stability}). The critical circle $X_1$ has radius $r=(a^{-1}-b^{-1})/2$, and it is centered on the real axis at $(-b^{-1}-r)$, where $a$ and $b$ are the lower and upper bounds of $1/RC(t)$. A very important feature of the system under study is dictated by the nature of the modulation function $C(t)$: With an external modulation or a properly filtered feedback, it always stays positive and bounded. This property ensures that  $a$ and $b$ are both positive. Therefore, circle $X_1$ lies entirely in the negative (left half) of the complex plane, and, consequently, it does not intersect the Nyquist locus of $G(s)$. Figure~\ref{fig4_Nyquist_stability}(b) schematically shows the  critical circle $X_1$ and the Nyquist locus of $G(s)$.

The stability of the circuit in Fig.~\ref{fig1_mimicking_arbitraryLandC}(c) can be also analyzed in a similar manner. In fact, it is possible to introduce the first-order differential equation for the charge on the capacitance as
\begin{equation}
\frac{\mathrm{d}q(t)}{\mathrm{d}t}+\frac{R_{\rm s}/R_C+1}{R_{\rm s}C(t)}q(t)=\frac{1}{R_{\rm s}}v_{\rm s}(t),
\end{equation}
which essentially has a form identical to Eq.~\eqref{eq_RCt_charge}. The Nyquist locus of $G(s)$ remains the same, and only function $F(s)$ changes. Practically, the coefficient in front of $q(t)$ is scaled, which only shifts bounds $a$ and $b$, while keeping them both positive. It means that circle $X_1$ is still lying entirely in the left half of the complex plane and never intersects $G(s)$.

Interestingly, the structures in Fig.~\ref{fig_invisible_sensor}(a--c) are also stable. An equivalent electric circuit for these devices is a parallel connection of a static capacitance with a time-varying one. Positive capacitance practically introduces a constant shift to the modulation function $C(t)$, which transforms Eq.~\eqref{eq_mod_function_for_C} as $C(t)=c_1/v_{\rm{cap}}(t)$. In a certain sense, this constant positive shift pushes the positive function $C(t)$ further from zero, which makes the system \textit{even more} stable.


\section{Conclusion}
In this work, we used the concept of temporal modulation and introduced a novel approach to realize static electromagnetic systems with arbitrary transfer functions. In particular, we showed how to properly modulate a capacitor in time to emulate the behavior of capacitance, inductance, and resistance having arbitrary values, including non-Foster reactive elements. We compared the theoretical results with the simulated ones with an excellent agreement between them confirming the theory presented in the paper. As an example of practical use, we proposed an idea of a planar invisible sensor, which senses the incident field and compensates for the scattering that it creates. Exploiting the analogy between a thin dielectric layer and a capacitive sheet, we suggested three different realization scenarios of the invisible sensor. Finally, in order to prove practical realizbaility of the presented approach, we studied stability of the proposed time-modulated structures by using the Nyquist stability criterion for dynamic systems. We showed that those structures remain stable under the assumption of external modulation, in contrast to potentially unstable conventional realizations of non-Foster circuits and materials. 


\section*{Acknowledgments}
This work was supported by the Academy of Finland under Grant No.~330260.


\bibliography{refs}
\bibliographystyle{IEEEtran}

\appendix

\section{Fields for alternative realizations of proposed invisible sensors}

\begin{figure}[!h]
    \centering
    \includegraphics[width=0.5\textwidth]{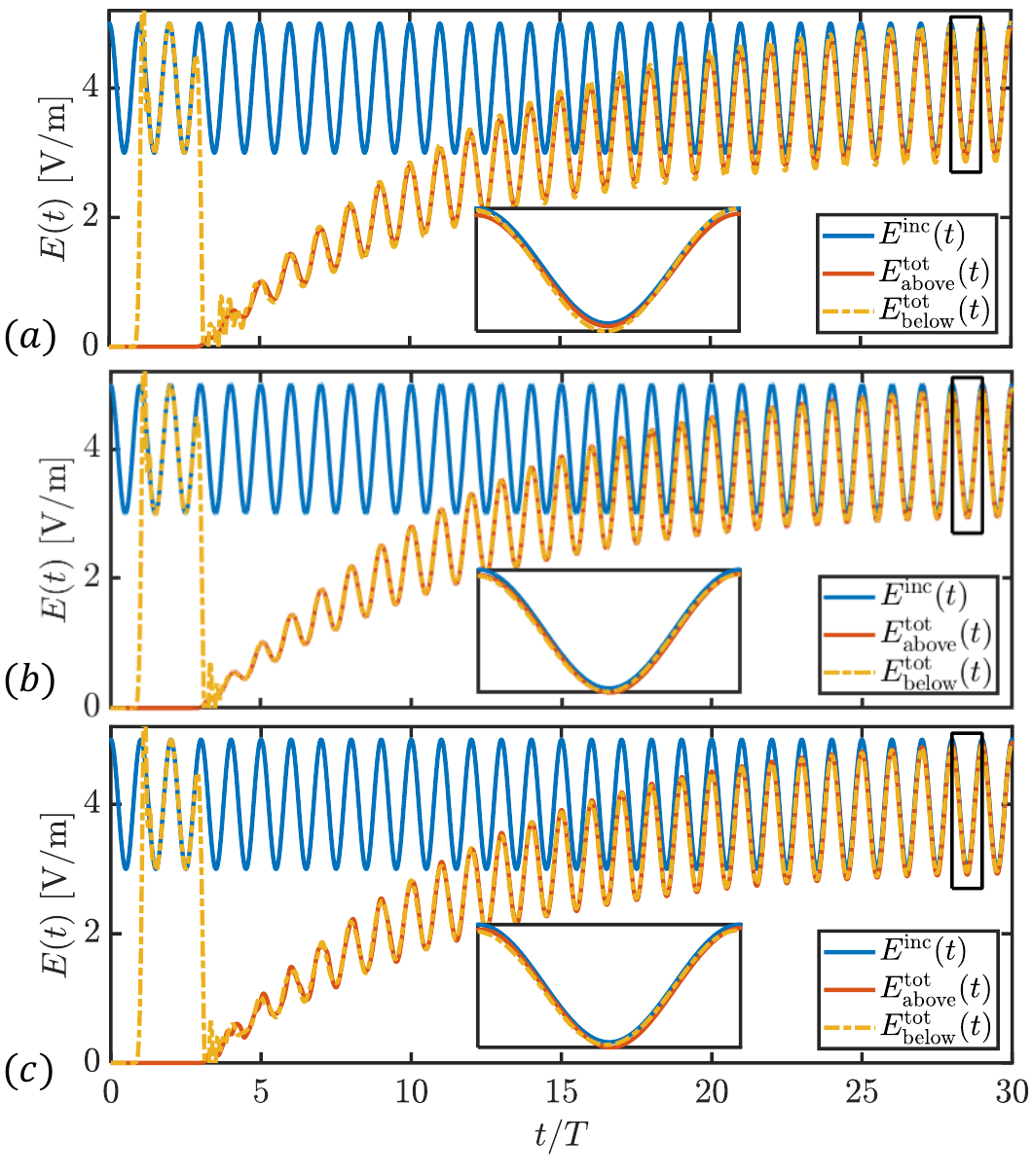}
    \caption{Total electric field above and below the metasurface compared with the incident field for three geometries proposed in main text. Panels (a),~(b), and~(c)  correspond to Figs.~\ref{fig_invisible_sensor}(a),~(b), and~(c), respectively.}
    \label{fig_suppl_Sens_iden_scat}
\end{figure}

Figure~\ref{fig_suppl_Sens_iden_scat} shows that all the realization designs produce almost identical scattering. Discrepancies emerging from the finite thickness of the capacitive layers remain negligibly small.

\section{Invisible devices with negligible absorption}

\begin{figure*}[!h]
    \centering
    \includegraphics[width=1\textwidth]{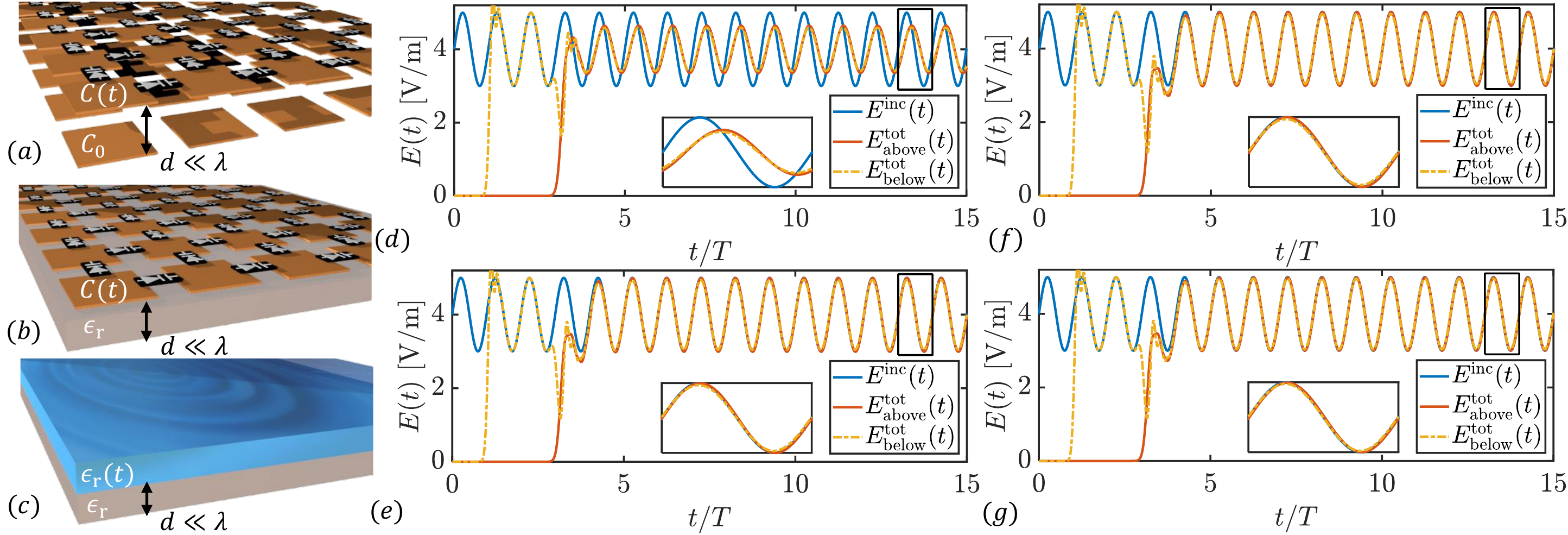}
    \caption{(a)--Concept of a broadband transparent metasurface made of two arrays of patches, separated by a small distance $d$. Capacitance of one the arrays is modulated in time using varactors. (b)--Conceptual realization of a transparent metasurface using an array of patches placed on top of a thin dielectric substrate. Varactors in the gaps between the patches realize variations of the layer parameters in time. (c)--Concept of a transparent metasurface formed by two thin dielectric layers. The permittivity of one of the layers is modulated in time. (d)~Total electric field above and below the metasurface compared with the incident field in case when the modulation is turned off. (e),~(f) and~(g)--Total electric field at the same points for conceptual realizations in (a),~(b) and~(c), respectively.}
    \label{fig_supl_not_a_Sensor}
\end{figure*}

Figures~\ref{fig_supl_not_a_Sensor}(a), (b) and~(c) show three possible realizations of an invisible layer with negligibly small resistance. These realizations are equivalent to those proposed in the main text, with the only difference in the resistive layer $R_0$. Here it is absent. All other parameters are assumed identical to those used in the main text. Figure~\ref{fig_supl_not_a_Sensor}(d) shows the total electric field above and below the metasurface at distance $\lambda$ compared to the incident field in the case when modulation is turned off. It is clear that the static capacitive layer creates strong scattering, meaning that the phase of the total field changes substantially. Figures~\ref{fig_supl_not_a_Sensor}(e), (f), and~(g) show the case when the modulation is turned on, and it clearly demonstrates how modulation makes the device  invisible, with the total fields above and below the metasurface identical to the incident field. Since all three designs are essentially a parallel connection of a capacitive layer and a time-varying one, scattered fields produced by them are identical.


\section{Power analysis}

\begin{figure}[!h]
    \centering
    \includegraphics[width=0.5\textwidth]{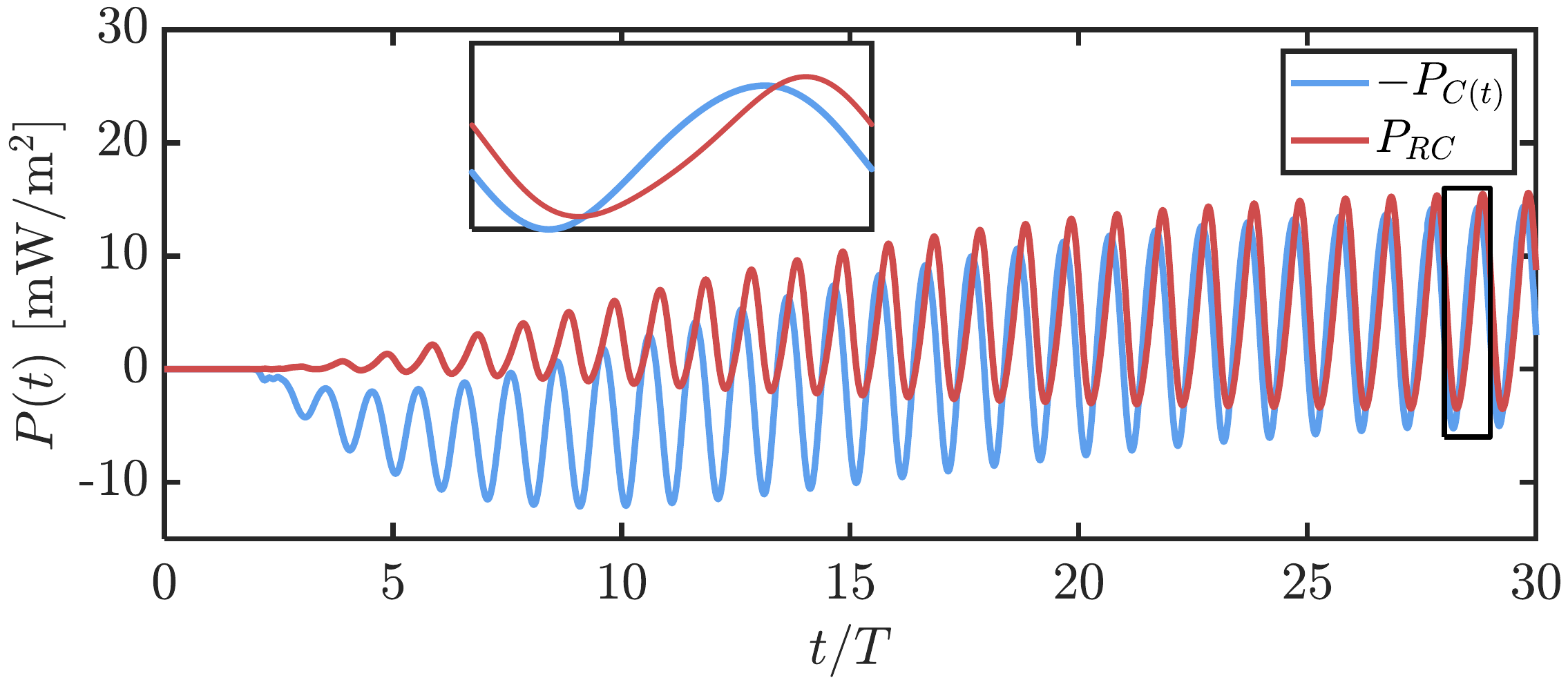}
    \caption{Power flowing through a static layer that is capacitive and resistive, compared to the power flowing through a time-varying capacitive layer. Note that the latter values are plotted with a  negative sign.}
    \label{fig_suppl_Powers}
\end{figure}

Here we investigate power distribution in time-varying structure shown in Fig.~\ref{fig_invisible_sensor}(a). The same analysis is applicable for structures in (b) and (c). It is interesting to see how the time-varying capacitive layer compensates the power absorbed by resistive layer. The red curve in Fig.~\ref{fig_suppl_Powers} shows the power that flows through static capacitive and resistive layers, and the blue curve shows the power that flows through the time-varying capacitive layer. The results show that overall the structure remains neutral, there is no loss and no gain, although a certain discrepancy is present. It can be explained by a non-zero thickness of the structure. Choosing a smaller separation distance between the layers, one could achieve a better matching. 

\end{document}